\documentclass[hyper]{JHEP} 

\usepackage{epsfig}

\newcommand\fverb{\setbox\pippobox=\hbox\bgroup\verb}

\newcommand\fverbdo{\egroup\medskip\noindent%

            \fbox{\unhbox\pippobox}\ }

\newcommand\fverbit{\egroup\item[\fbox{\unhbox\pippobox}]}

\newbox\pippobox

\title{$(m,n)-$String and D1-Brane in Stringy  Newton-Cartan Background }
\author{J. Kluso\v{n}\\
Department of
Theoretical Physics and Astrophysics\\
Faculty of Science, Masaryk University\\
Kotl\'{a}\v{r}sk\'{a} 2, 611 37, Brno\\
Czech Republic\\
E-mail: \email{klu@physics.muni.cz}} \preprint{}

 \abstract{This paper is devoted to the analysis of $(m,n)-$ string in stringy Newton-Cartan background. We start with the Hamiltonian constraint for $(m,n)-$string in general background and perform limiting procedure on metric and     NSNS and Ramond-Ramond two form background that leads to stringy Newton-Cartan gravity. We also analyze conditions
    that these background fields have to obey in order to define
    consistent world-sheet $(m,n)-$theory. We also discuss D1-brane with dynamical
    electric field in stringy Newton-Cartan gravity.}

\def\he{\hat{e}}

\def\tn{\tilde{n}}

\def\bA{\mathbf{A}}

\def\bB{\mathbf{B}}

\def\be{\begin{equation}}

\def\ee{\end{equation}}

\def\bea{\begin{eqnarray}}
\def\bh{\bar{h}}
\def\eea{\end{eqnarray}}

\def\bX{\mathbf{X}}
\def\bY{\mathbf{Y}}

\def\mH{\mathcal{H}}

\def\tC{\tilde{C}}

\newcommand{\tchi}{\tilde{\chi}}

\newcommand{\hh}{\hat{h}}

\newcommand{\mF}{\mathcal{F}}

\newcommand{\mG}{\mathcal{G}}

\def \bA{\mathbf{A}}

\newcommand{\ba}{\mathbf{a}}

\newcommand{\mL}{\mathcal{L}}

\def\pb #1{\left\{#1\right\}}

\def\htau{\hat{\tau}}
\begin{document}
\section{Introduction}
Recently new interesting generalization of Newton-Cartan (NC) gravity \cite{Cartan:1923zea}
 was proposed in \cite{Andringa:2012uz}. In standard NC gravity there is one-dimensional foliation direction of space-time corresponding to the absolute time direction that is longitudinal to the world-line of particle. The stringy NC gravity is generalization of this picture in the sense that one dimensional
 foliation is replaced by a two-dimensional foliation with one time-like and spatial foliation directions that are longitudinal  to the world-sheet of the string. Further, stringy NC gravity is related to non-relativistic strings
\cite{Gomis:2000bd,Danielsson:2000gi} in the same way as general
relativity is related to relativistic string theory as was recently
discussed in \cite{Bergshoeff:2018yvt} \footnote{For alternative
definition of non-relativistic strings with $AdS/CFT$ applications,
see \cite{Harmark:2018cdl,Harmark:2017rpg}. The Hamiltonian analysis
of this model was performed in \cite{Kluson:2018egd}.}. This paper
also analyzed T-duality properties of non-relativistic strings in
stringy Newton-Cartan background that are very interesting when it
was shown that
 T-duality along the longitudinal direction of the stringy
  Newton-Cartan geometry describes relativistic string theory on a Lorentzian geometry with a compact lightlike isometry, which is otherwise only defined by a subtle infinite boost limit. This fact was further confirmed in
\cite{Kluson:2018vfd} when T-duality properties of non-relativistic string in
stringy NC background was analyzed with the help of canonical formulation of this string sigma model  found in \cite{Kluson:2018grx}.

 Since the proposal suggested in \cite{Bergshoeff:2018yvt} is very interesting we mean that it is natural to study further aspects of  stringy NC. In particular, we would like to see whether it is possible to define another extended objects known in string theory, as for example D-branes \cite{Polchinski:1995mt} in this background. It is natural to begin with  D1-brane which is two dimensional object with
 gauge field propagating on its world-sheet. This action couples to Ramond-Ramond (RR) two form and one form together to gravity, NSNS two form and dilation that are  background fields of type IIB supergravity. Such an action can be written in  manifestly $SL(2,Z)$ invariant form \cite{Cederwall:1997ts,Townsend:1997kr} that reflects $SL(2,Z)$ duality of type IIB theory. We consider Hamiltonian for this object and analyze non-relativistic limit of the background metric as was proposed in
 \cite{Bergshoeff:2015uaa}
 and that was used for the definition of non-relativistic string in stringy NC background in   \cite{Kluson:2018uss}. In case of D1-brane as a probe the situation is more interesting since there is an electric flux on the world-sheet of D1-brane. We firstly  consider D1-brane with fixed gauge invariance when the momentum conjugate to $A_\sigma$ is equal to some integer number $m$ that counts the number of fundamental strings in the bound state with D1-brane. We call resulting object as $(m,n)-$string \cite{Cederwall:1997ts,Townsend:1997kr}. If we start with such a probe we should define background RR and NSNS two form fields in such a way that the limiting procedure \cite{Bergshoeff:2015uaa} leads to finite Hamiltonian constraint. It turns out that this procedure is similar to the case of fundamental string
 \cite{Kluson:2018uss}. However due to the non-trivial structure of the background fields that define  stringy Newton-Cartan geometry we have to check that they solve the background equations of motion of type IIB gravity. More explicitly, we consider type IIB gravity equations of motion and study their solutions for the metric ansatz proposed in  \cite{Bergshoeff:2015uaa} together with  another fields that appear in  $(m,n)-$ string action.. Inserting the ansatz \cite{Bergshoeff:2015uaa} into definition of the relativistic Christoffel symbols we derive condition when this connection is finite even if the parameter that defines non-relativistic limit goes to infinity. We restrict ourselves to the case of zero torsion condition
\cite{Bergshoeff:2017dqq} even if certainly more general situations are possible, see for example \cite{VandenBleeken:2017rij}. Now since the Christoffel symbols is finite we also find that Ricci tensor is finite and hence in order to have well defined non-relativistic limit in the background equations of motion we have to demand that the stress energy tensor for the background fields is finite as well. It turns out that this condition has important consequence on the dilaton and RR zero form which now have to be constant. Then the equation of motion for dilaton RR and zero form implies that field strengths of RR and NSNS two form fields have to vanish. We also discuss more interesting case of the fundamental string when
now dilaton does not have to be constant. In this case however we find that requirement that Christoffel symbols for Einstein frame metric is finite implies that
 spatial projection of derivative of dilaton should be zero.

In the second part of the paper we focus on the problem of the definition of the action for  $n$ coincident D1-branes in the stringy NC background when the gauge symmetry on the world-sheet is unfixed. Then in order to have finite and non-trivial theory we have to scale RR two form and zero form field in appropriate way as well. As a result we obtain finite Hamiltonian constraint for D1-brane in stringy NC background. Then following standard procedure we find corresponding Lagrangian density. Finally we also discuss the equations of motion for the background fields and we find that they have to obey the same conditions as in case of the stringy NC gravity.

This paper is organized as follows. In the next section (\ref{second}) we introduce an action and Hamiltonian for $n$ coincident D1-branes in general background. Then in section (\ref{third}) we consider gauge fixed theory that corresponds to $(m,n)-$ string and find its non-relativistic action in NC background.  In section (\ref{fourth}) we analyze equations of motion for background fields. In section (\ref{fifth}) we study D1-brane in stringy Newton-Cartan background when the gauge symmetry is not fixed and we find corresponding Lagrangian density. Finally in conclusion (\ref{sixth}) we outline our results and suggest possible extension of this work.
\section{Hamiltonian Formulation of D1-brane in General Background }\label{second}
In this section we review basic facts about D1-brane in general background and its Hamiltonian formulation.
As the starting point we consider an action for
 $n$ coincident D1-branes in
general background
\begin{eqnarray}\label{D1branegen}
& &S=-nT_{D1}\int d\tau d\sigma e^{-\Phi}\sqrt{-\det
    \bA}+\nonumber \\
& &+nT_{D1}\int d\tau d\sigma
((b_{\tau\sigma}+2\pi\alpha'\mF_{\tau\sigma})\chi+ c_{\tau\sigma}) \
,
\nonumber \\
& &\bA_{\alpha\beta}=G_{\mu\nu}\partial_\alpha x^\mu
\partial_\beta x^\nu+2\pi\alpha'\mF_{\alpha\beta}+B_{\mu\nu}
\partial_\alpha x^\mu\partial_\beta x^\nu \ , \nonumber \\
& &\mF_{\alpha\beta}=\partial_\alpha A_\beta-\partial_\beta A_\alpha
\
, \nonumber \\
\end{eqnarray}
where $x^\mu,\nu=0,1,\dots,9$ are embedding coordinates of D1-brane in
the background that is specified by the metric $G_{\mu\nu}$ and NSNS two
form  $B_{\mu\nu}=-B_{\nu\mu}$ together with Ramond-Ramond two form
$C^{(2)}_{\mu\nu}=-C^{(2)}_{\nu\mu}$.  There are also two background scalar fields,  dilaton $\Phi$ and RR zero form $\chi$. Further,
$\sigma^\alpha=(\tau,\sigma)$ are world-sheet coordinates  and
$b_{\tau\sigma},c_{\tau\sigma}$ are pull-backs of $B_{\mu\nu}$ and
$C_{\mu\nu}^{(2)}$ to the world-sheet of D1-brane. Explicitly,
\begin{equation}
b_{\alpha\beta}= B_{\mu\nu}\partial_\alpha x^\mu\partial_\beta x^\nu \
, \quad  c_{\alpha\beta}=C^{(2)}_{\mu\nu}\partial_\alpha x^\mu\partial_\beta
x^\nu  \ .
\end{equation}
Finally $T_{D1}=\frac{1}{2\pi\alpha'}$ is D1-brane tension and
$A_{\alpha},\alpha=\tau,\sigma$ is two dimensional gauge field that
propagates on the world-sheet of D1-brane.

It is useful to rewrite the action (\ref{D1branegen}) into the form
\begin{eqnarray}\label{D1actional}
S&=&-nT_{D1}\int d\tau d\sigma e^{-\Phi} \sqrt{-\det g-
    (2\pi\alpha'\mF_{\tau\sigma}+b_{\tau\sigma})^2} \nonumber \\
&+&nT_{D1}\int d\tau d\sigma
((b_{\tau\sigma}+2\pi\alpha'\mF_{\tau\sigma})\chi+ c_{\tau\sigma})
\ , \nonumber \\
\end{eqnarray}
where $g_{\alpha\beta}=G_{\mu\nu}\partial_\alpha x^\mu
\partial_\beta x^\nu, \det g=g_{\tau\tau}g_{\sigma\sigma}-
(g_{\tau\sigma})^2$.
Now we  proceed  to the Hamiltonian formulation of the theory
defined by the action (\ref{D1actional}). First of all we derive
conjugate momenta to $x^\mu$ and $A_\alpha$ from (\ref{D1actional})
\begin{eqnarray}\label{defpM}
& &p_\mu=\frac{\delta L}{\delta \partial_\tau x^\mu}= nT_{D1}
\frac{e^{-\Phi}}{\sqrt{-\det g -(2\pi\alpha'F_{\tau\sigma}+
        b_{\tau\sigma})^2}}(G_{\mu\nu}\partial_\alpha x^\nu g^{\alpha \tau}\det g+\nonumber \\
& &+(2\pi\alpha'F_{\tau\sigma}+ b_{\tau\sigma})B_{\mu\nu}\partial_\sigma
x^\nu)+nT_{D1}(\chi B_{\mu\nu}\partial_\sigma x^\nu+C^{(2)}_{\mu\nu}\partial_\sigma x^\nu) \ , \nonumber \\
& &\pi^\sigma=\frac{\delta L}{\delta
    \partial_\tau A_\sigma}=\frac{ne^{-\Phi}(2\pi\alpha'F_{\tau\sigma}+
    b_{\tau\sigma})}{\sqrt{-\det g -(2\pi\alpha'F_{\tau\sigma}+
        b_{\tau\sigma})^2}}+n\chi\ , \quad \pi^\tau=\frac{\delta L}{\delta
    \partial_\tau A_
    \tau}\approx 0 \nonumber \\
\end{eqnarray}
and hence
\begin{eqnarray}
\Pi_\mu&\equiv&
p_\mu-T_{D1}\pi^\sigma B_{\mu\nu}\partial_\sigma x^\nu-nT_{D1}C^{(2)}_{\mu\nu}\partial_\sigma
x^\nu= \nonumber \\
&=&nT_{D1} \frac{e^{-\Phi}}{\sqrt{-\det g
        -(2\pi\alpha'F_{\tau\sigma}+
        b_{\tau\sigma})^2}}G_{\mu\nu}\partial_\alpha x^\nu g^{\alpha \tau}\det g \
.
\nonumber \\
\end{eqnarray}
Using these relations it is easy to see that the bare Hamiltonian is
equal to
\begin{eqnarray}\label{Hgen}
H_B=\int d\sigma(p_\mu\partial_\tau x^\mu+\pi^\sigma \partial_\tau
A_\sigma-\mL)= \int d\sigma \pi^\sigma\partial_\sigma A_\tau
\nonumber \\
\end{eqnarray}
while we have three primary constraints
\begin{eqnarray}\label{HD1gen}
& &\pi^\tau\approx 0 \ , \quad \mH_\sigma\equiv p_\mu\partial_\sigma x^\mu\approx 0 \ , \nonumber \\
& &\mH_\tau\equiv \Pi_\mu G^{\mu\nu}\Pi_\nu+
T^2_{D1}\left(n^2e^{-2\Phi}+\left(\pi^\sigma -n\chi\right)^2
\right)g_{\sigma\sigma}\approx 0 \ .
\end{eqnarray}
Including
these primary constraints to the definition of the Hamiltonian we
obtain an extended Hamiltonian in the form
\begin{equation}
H=\int d\sigma (N^\tau\mH_\tau+N^\sigma
\mH_\sigma-A_\tau\partial_\sigma\pi^\sigma+v_\tau \pi^\tau) \ ,
\end{equation}
where $N^\tau,N^\sigma,v_\tau$ are Lagrange multipliers
corresponding to the primary constraints $\mH_\tau\approx 0 \
,\mH_\sigma\approx 0 \ , \pi^\tau\approx 0$.
Now we have to check the stability of all constraints. The
requirement of the preservation of the primary constraint
$\pi^\tau\approx 0$ implies the secondary constraint
\begin{equation}
\mG=\partial_\sigma \pi^\sigma\approx 0 \ .
\end{equation}
Now we are ready to proceed to the analysis of $(m,n)-$string in stringy
NC background.
\section{$(m,n)-$String in Stringy NC Background}\label{third}
$(m,n)-$ string is defined as D1-brane where the gauge symmetry is fixed
so that the electric flux $\pi^\sigma$ is constant and counts the number
of fundamental strings. Explicitly,  we  fix the gauge
generated by $\mG$ with the gauge fixing function
$A_\sigma=\mathrm{const}$. Then  the fixing of the gauge implies
that $\pi^\sigma=f(\tau)$ but the equation of motion for
$\pi^\sigma$ implies that $\partial_\tau \pi^\sigma=0$ and hence
$\pi^\sigma=m$, where $m$ is integer that counts the number of
fundamental strings bound to $n$ D1-branes.

In order to define stringy NC background we follow \cite{Bergshoeff:2015uaa}
and  introduce $10-$ dimensional vierbein
 $E_\mu^{ \ A}$ so that the metric components have the form
\begin{equation}
G_{\mu\nu}=E_\mu^{ \ A}E_\nu^{ \ B}\eta_{AB} \ , \eta_{AB}=\mathrm{diag}(-1,\dots,1) \ .
\end{equation}
Note that the metric inverse $G^{\mu\nu}$ is defined with the help of the inverse vierbein $E^\mu_{ \ B}$ that obeys the relation
\begin{equation}
E_\mu^{ \ A}E^\mu_{ \ B}=\delta^A_{B} \  ,  \quad E_\mu^{ \ A}E^\nu_{ \ A}=
\delta^\mu_{\nu} \ .
\end{equation}
As the next step
we  split target-space indices $A$ into $A=(a',a)$ where now $a'=0,1$ and $a=2,\dots,d-1$. Then we introduce  longitudinal vielbein $\tau_\mu^{ \ a}$ so that we write
\begin{equation}
  \tau_{\mu\nu}=\tau_\mu^{ \ a}\tau_\nu^{ \ b}
\eta_{ab} \ , \quad a,b=0,1 \ .
\end{equation}
In the same way we introduce vierbein $e_\mu^{ \ a'}, a=2,\dots,d-1$ and also  introduce gauge field  $m_\mu^{ \ a}$. The $\tau_\mu^{ \ a}$ can be interpreted as the gauge fields of the longitudinal translations while $e_\mu^{ \ a'}$  as the gauge fields of the transverse translations
\cite{Andringa:2012uz}. Then we can also introduce their inverses with respect to their longitudinal and transverse translations
\begin{eqnarray}
e_\mu^{ \ a'}e^\mu_{ \ b'}=\delta^{a'}_{b'} \ ,  \quad
e_\mu^{ \ a'}e^\nu_{ \ a'}=\delta_\mu^\nu-\tau_\mu^{ \ a}
\tau^\nu_{ \ a} \ , \quad \tau^\mu_{ \ a}\tau_\mu^{ \ b}=\delta_a^b \ , \quad
\tau^\mu_{ \ a}e_\mu^{ \ a'}=0 \ , \quad
\tau_\mu^{ \ a}e^\mu_{ \ a'}=0 \ . \nonumber \\
\end{eqnarray}
With the help of these fields  we introduce following
parametrization of relativistic vierbein \cite{Bergshoeff:2015uaa}
\begin{equation}\label{relvierin}
E_\mu^{ \ a}=\omega \tau_\mu^{ \ a}+\frac{1}{2\omega}m_\mu^{ \ a} \ , \quad
E_\mu^{ \ a'} =e_\mu^{ \ a'} \ ,
\end{equation}
where $\omega$ is free parameter that we take to infinity when we define non-relativistic limit. Note that the inverse vierbein to (\ref{relvierin}) has the form  (up to terms of order $\omega^{-3}$)
\begin{equation}\label{relvierinv1}
E^\mu_{ \ a}=\frac{1}{\omega}\tau^\mu_{ \ a}-\frac{1}{2\omega^3}\tau^\mu_{ \ b}m_\rho^{ \ b}
\tau^\rho_{ \ a} \ , \quad E^\mu_{ \ a'}=e^\mu_{ \ a'}
-\frac{1}{2\omega^2} \tau^\mu_{ \ a}m_\rho^{ \ a}e^\rho_{ \ a'} \ .
\end{equation}
Then with the help of (\ref{relvierin}) and (\ref{relvierinv1}) we obtain following form of the metric
\begin{eqnarray}\label{metNCans}
G_{\mu\nu}
&=&\omega^2 \tau_{\mu\nu}+h_{\mu\nu}+\frac{1}{2}\tau_\mu^{ \ a}m_\nu^{ \ b}\eta_{ab}+
\frac{1}{2}m_\mu^{ \ a}\tau_\nu^{ \ b}\eta_{ab}+\frac{1}{4\omega^2}m_\mu^{ \ a}m_\nu^{ \ b}
\eta_{ab} \ , \nonumber \\
G^{\mu\nu}
&=& \frac{1}{\omega^2}\tau^{\mu\nu}+h^{\mu\nu}
-\frac{1}{2\omega^2}(\tau^\nu_{ \ b}m_\rho^{ \ b}h^{\rho\mu}
+\tau^\mu_{ \ b}m_\rho^{ \ b}h^{\rho\nu})
-\nonumber \\
&-&\frac{1}{2\omega^4}
(\tau^\mu_{ \ c}m^c_{ \ \rho}\tau^{\rho\nu}+
\tau^\nu_{ \ d}m^d_{ \ \rho}\tau^{\rho\mu})
+\frac{1}{4\omega^4}\tau^\mu_{ \ a}m_\rho^{ \ a}
h^{\rho\sigma}\tau^\nu_{ \ b}m_\sigma^{\ b} \ . \nonumber \\
\end{eqnarray}
As the next step we have to introduce an appropriate parametrization
of RR and  NSNS two forms. Following analysis presented
\cite{Kluson:2017abm} we suggest that they have the form
\begin{eqnarray}\label{ansB}
B_{\mu\nu}&=&
\bX(\omega\tau_\mu^{ \ a}-\frac{1}{2\omega}m_\mu^{ \ a})(\omega \tau_\nu^{ \ b}-\frac{1}{2\omega}m_\nu^{ \ b})\epsilon_{ab}+b_{\mu\nu}
=\nonumber \\
&=&\bX(\omega^2\tau_\mu^{ \ a}\tau_\nu^{ \ b}\epsilon_{ab}-
\frac{1}{2}(m_\mu^{ \ a}\tau_{\nu}^{ \ b}+
\tau_\mu^{\  a}m_\nu^{ \ b})\epsilon_{ab}+\frac{1}{4\omega^2}
m_\mu^{ \ a}m_\mu^{ \ b}\epsilon_{ab})+b_{\mu\nu} \ , \nonumber \\
C_{\mu\nu}&=&
\bY(\omega\tau_\mu^{ \ a}-\frac{1}{2\omega}m_\mu^{ \ a})(\omega \tau_\nu^{ \ b}-\frac{1}{2\omega}m_\nu^{ \ b})\epsilon_{ab}+b_{\mu\nu}
=\nonumber \\
&=&\bY(\omega^2\tau_\mu^{ \ a}\tau_\nu^{ \ b}\epsilon_{ab}-
\frac{1}{2}(m_\mu^{ \ a}\tau_{\nu}^{ \ b}+
\tau_\mu^{\  a}m_\nu^{ \ b})\epsilon_{ab}+\frac{1}{4\omega^2}
m_\mu^{ \ a}m_\mu^{ \ b}\epsilon_{ab})+c_{\mu\nu} \ , \nonumber \\
\nonumber \\
\end{eqnarray}
where
\begin{equation}
\epsilon_{ab}=-\epsilon_{ba} \ , \quad  \epsilon_{01}=1 \ ,
\nonumber \\
\end{equation}
and where $\bX,\bY$ are factors that can depend on space-time fields and $m,n$ that
we choose in such a way to make Hamiltonian constraint finite in the limit $\omega\rightarrow \infty$. In fact it turns out that  this requirement implies
\begin{equation}\label{result}
m\bX+n\bY=\sqrt{n^2 e^{-2\Phi}+(m-n\chi)^2
} \ .
\end{equation}
Then using (\ref{ansB}) and also (\ref{result})  we can now perform the limit $\omega\rightarrow \infty$ in the Hamiltonian constraint and we find that it is equal to
\begin{eqnarray}
& &\mH_\tau=\Pi_\mu h^{\mu\nu}\Pi_\nu-
2T_{D1}\sqrt{n^2 e^{-2\Phi}+(m-n\chi)^2}
\Pi_\mu \htau^\mu_{ \ a}\eta^{ab}\epsilon_{bc}\tau_\rho^{ \ c}
\partial_\sigma x^\rho +\nonumber \\
& &+T_{D1}^2 (n^2 e^{-2\Phi}+(m-n\chi)^2)\partial_\sigma x^\mu \bar{h}_{\mu\nu}\partial_\sigma x^\nu-\nonumber \\
& &-T_{D1}^2(n^2 e^{-2\Phi}+(m-n\chi)^2)\partial_\sigma x^\nu \tau_\nu^{ \ c} \epsilon_{cd}
\Phi^{da}\epsilon_{ab}\tau_\nu^{ \ f}\partial_\sigma x^\nu \ ,
\nonumber \\
\end{eqnarray}
where we introduced following fields
\begin{eqnarray}
& &\htau^\mu_{ \ a}=\tau^\mu_{ \ a}-h^{\mu\nu}m_\nu^{ \ b}\eta_{ba} \ , \quad
\he_\mu^{ \ a'}=e_\mu^{ \ a'}+m_\nu^{ \ a}e^\nu_{ \ c'}\delta^{c'a'}
\tau_\mu^{ \ b}\eta_{ba} \ , \nonumber \\
& &\Phi^{ab}=-\tau^\mu_{ \ d}\eta^{da}m_\mu^{ \ b}-m_\mu^{ \ a}\tau^\mu_{ \ d}
\eta^{db}+m_\mu^{ \ a}h^{\mu\nu}m_\nu^{ \ b} \ , \nonumber \\
& & \bar{h}_{\mu\nu}=h_{\mu\nu}+m_\mu^{ \ a}\tau_\nu^{ \ b}\eta_{ab}+
\tau_\mu^{ \ a}m_\nu^{ \ b}\eta_{ab} \ , \nonumber \\
\end{eqnarray}
and where
\begin{equation}
\Pi_\mu=p_\mu-T_{D1}mb_{\mu\rho}\partial_\sigma x^\rho-
T_{D1}nc_{\mu\rho}\partial_\sigma x^\rho \ .
\end{equation}
Finally using the fact that
\begin{eqnarray}
& &-\partial_\sigma x^\mu \tau_\mu^{ \ a}\epsilon_{ab}
\Phi^{bc}\epsilon_{cd}\tau_\nu^{ \ d}\partial_\sigma x^\nu=
\tau^{ \ c}_\sigma\eta_{ca}\Phi^{ab}\eta_{bd}\tau^{ \ d}_\sigma-\ba_{\sigma\sigma}
\Phi^{ab}\eta_{ba} \ , \nonumber \\
& & \tau^{ \ a}_\alpha=\tau_\mu^{ \ a}\partial_\alpha x^\mu \ , \quad
\ba_{\alpha\beta}=\partial_\alpha x^\mu \tau_{\mu\nu}\partial_\beta x^\nu
\nonumber \\
\end{eqnarray}
we can rewrite the Hamiltonian constraint into the form
\begin{eqnarray}\label{mHmn}
& &\mH_\tau=\Pi_\mu h^{\mu\nu}\Pi_\nu-
2T_{D1}\sqrt{n^2 e^{-2\Phi}+(m-n\chi)^2}
\Pi_\mu \htau^\mu_{ \ a}\eta^{ab}\epsilon_{bc}\tau_\rho^{ \ c}
\partial_\sigma x^\rho +\nonumber \\
& &+T_{D1}^2 (n^2 e^{-2\Phi}+(m-n\chi)^2)\partial_\sigma x^\mu \bar{h}_{\mu\nu}\partial_\sigma x^\nu+\nonumber \\
& &+T_{D1}^2(n^2 e^{-2\Phi}+(m-n\chi)^2)\left(\tau^{ \ c}_\sigma\eta_{ca}\Phi^{ab}\eta_{bd}\tau^{ \ d}_\sigma-\ba_{\sigma\sigma}
\Phi^{ab}\eta_{ba}\right) \ .
\nonumber \\
\end{eqnarray}
This is the Hamiltonian constraint for $(m,n)-$ string in stringy NC background.

Let us now proceed to the Lagrangian form of this theory. Note that the Hamiltonian
is equal to
\begin{equation}\label{Hmn}
H=\int d\sigma (N^\tau\mH_\tau+N^\sigma \mH_\sigma) \ ,
\end{equation}
where $\mH_\tau$ is given in (\ref{mHmn}). From (\ref{Hmn}) we obtain
\begin{eqnarray}
\dot{x}^\mu=\pb{x^\mu,H}=
2N^\tau h^{\mu\nu}\Pi_\nu-2N^\tau T_{D1}
\sqrt{n^2 e^{-2\Phi}+(m-n\chi)^2}\htau^\mu_{ \ a}\eta^{ab}\epsilon_{bc}
\tau_\rho^{ \ c}\partial_\sigma x^\rho
+N^\sigma \partial_\sigma x^\mu \ ,
\nonumber \\
\end{eqnarray}
where $\dot{x}^\mu\equiv \partial_\tau x^\mu \ , x'^\mu\equiv
\partial_\sigma x^\mu$.
To proceed further we use the fact that
\begin{equation}
\he_\mu^{ \ a'}\tau^\mu_{ \ b}=0 \ , \quad \he_\mu^{ \ a'}h^{\mu\nu}=e^\nu_{ \ b'}\delta^{b' a'}
\end{equation}
and hence
\begin{eqnarray}
(\dot{x}^\mu-N^\sigma x'^\mu)
\he_{\mu}^{ \ a'}\delta_{a' b'}
\he_{\nu}^{ \ b'  }(\dot{x}^\nu-N^\sigma x'^\nu)=4(N^\tau)^2 \Pi_\mu h^{\mu\nu}\Pi_\nu
\ .
\nonumber \\
\end{eqnarray}
Then the Lagrangian density has the form
\begin{eqnarray}\label{mnL}
& &\mL=\dot{x}^\mu p_\mu-\mH=(\dot{x}^\mu-N^\sigma x'^\mu)p_\mu-N\mH_\tau=
\nonumber \\
& &=\frac{1}{4N^\tau}(\dot{x}^\mu-N^\sigma x'^\mu)
\he_{\mu}^{ \ a'}\delta_{a' b'}
\he_{\nu}^{ \ b'  }(\dot{x}^\nu-N^\sigma x'^\nu)-\nonumber \\
& &-T_{D1}^2N^\tau (n^2 e^{-2\Phi}+(m-n\chi)^2)\partial_\sigma x^\mu \bar{h}_{\mu\nu}\partial_\sigma x^\nu-\nonumber \\
& &-N^\tau T_{D1}^2(n^2 e^{-2\Phi}+(m-n\chi)^2)\left(\tau^{ \ c}_\sigma\eta_{ca}\Phi^{ab}\eta_{bd}\tau^{ \ d}_\sigma-\ba_{\sigma\sigma}
\Phi^{ab}\eta_{ba}\right)
\nonumber \\
& &+T_{D1}(m\dot{x}^\mu b_{\mu\nu}x'^\nu+n\dot{x}^\mu c_{\mu \nu}x'^\nu) \ .
\nonumber \\
\end{eqnarray}
To proceed further we use the fact that
\begin{equation}
\he_\mu^{ \ a'}\delta_{a'b'}\he_\nu^{ \ b'}=\bh_{\mu\nu}+
\tau_\mu^{ \ c}\eta_{ca}\Phi^{ab}\eta_{bd}\tau_\nu^{ \ d} \ , \quad
\bh_{\mu\nu}=h_{\mu\nu}+m_\mu^{ \ a}\tau_\nu^{ \ b}\eta_{ab}+
m_\nu^{ \ a}\tau_\mu^{ \ b}\eta_{ab} \ .
\end{equation}
Further, the special property of non-relativistic string and $(m,n)-$string  is
that $N^\tau,N^\sigma$  are determined by the equations of motion for $x^\mu$. To see this, we multiply
the equation of motion for $x^\mu$ by $\tau_{\mu\nu}$ and we obtain
\begin{equation}
\tau_{\mu\nu}(\dot{x}^\nu-N^\sigma x'^\nu)=
-2N^\tau T_{D1}
\sqrt{n^2 e^{-2\Phi}+(m-n\chi)^2}\tau_{\mu\nu}\htau^\nu_{ \ a}\eta^{ab}\epsilon_{bc}
\tau_\rho^{ \ c}\partial_\sigma x^\rho \ .
\end{equation}
If we multiply this equation with $\partial_\sigma x^\mu$ we get
\begin{equation}\label{Nsigmasol}
N^\sigma=\frac{\ba_{\tau\sigma}}{\ba_{\sigma\sigma}} \ .
\end{equation}
In the similar way we obtain
\begin{eqnarray}
(\partial_\tau x^\mu -N^\sigma \partial_\sigma x^\mu)\tau_{\mu\nu}
(\partial_\tau x^\nu-N^\sigma \partial_\sigma x^\nu)
=-4(N^\tau)^2T_{D1}^2
(n^2 e^{-2\Phi}+(m-n\chi)^2)\ba_{\sigma\sigma} \  \nonumber \\
\end{eqnarray}
that can be solved for $N^\tau$ as
\begin{equation}\label{Ntausol}
N^\tau=\frac{1}{2\ba_{\sigma\sigma}\sqrt{n^2 e^{-2\Phi}+(m-n\chi)^2}}\sqrt{-\det\ba} \ .
\end{equation}
Inserting (\ref{Nsigmasol}) and (\ref{Ntausol}) into (\ref{mnL})
 we obtain Lagrangian density in the form
\begin{eqnarray}\label{mnfinal}
& &\mL=-\frac{1}{2}\sqrt{n^2 e^{-2\Phi}+(m-n\chi)^2}\sqrt{-\det\ba}
\ba^{\alpha\beta}\bh_{\alpha\beta}+\nonumber \\
& & +T_{D1}(m\dot{x}^\mu b_{\mu\nu}x'^\nu+n\dot{x}^\mu c_{\mu \nu}x'^\nu) \ ,
\nonumber \\
\end{eqnarray}
where we used the result found in
  \cite{Kluson:2018uss}
that  two terms proportional to $\Phi^{ab}$ cancel each  other as
\begin{equation}
\ba^{\alpha\beta}\tau_\alpha^{ \ a}\eta_{ab}\Phi^{bc}\eta_{cd}\tau_\beta^{ \ d}-
\Phi^{ab}\eta_{ba}
=0 \ .
\end{equation}
%
The Lagrangian density (\ref{mnfinal}) is  the final form of  the Lagrangian density for $(m,n)-$string
in  stringy NC background. However in order to be our approach self-consistent we have to check  the equations of motion for the background fields that were used in the construction of the Lagrangian density (\ref{mnfinal}).

\section{Background Equation of Motion}\label{fourth}
In order to derive form of the non-relativistic $(m,n)-$string we presumed
special form of the background fields. Our goal is to check the consistency of
the equations of motion for the background fields when the limiting procedure is performed.

We start with  Christoffel symbols of parent relativistic theory
\begin{equation}\label{defGamma}
\Gamma_{\mu\nu}^{\rho}=\frac{1}{2}G^{\rho\sigma}(\partial_\mu G_{\sigma\nu}+
\partial_\nu G_{\sigma\mu}-
\partial_\sigma G_{\mu\nu}) \ .
\end{equation}
Now recall that the metric components that define stringy NC background
  have the form
\begin{eqnarray}\label{metback}
G_{\mu\nu}
&=&\omega^2 \tau_{\mu\nu}+h_{\mu\nu}+\frac{1}{2}\tau_\mu^{ \ a}m_\nu^{ \ b}\eta_{ab}+
\frac{1}{2}m_\mu^{ \ a}\tau_\nu^{ \ b}\eta_{ab}+\frac{1}{4\omega^2}m_\mu^{ \ a}m_\nu^{ \ b}
\eta_{ab} \ , \nonumber \\
G^{\mu\nu}
&=& \frac{1}{\omega^2}\tau^{\mu\nu}+h^{\mu\nu}
-\frac{1}{2\omega^2}(\tau^\nu_{ \ b}m_\rho^{ \ b}h^{\rho\mu}
+\tau^\mu_{ \ b}m_\rho^{ \ b}h^{\rho\nu})
-\nonumber \\
&-&\frac{1}{2\omega^4}
(\tau^\mu_{ \ c}m^c_{ \ \rho}\tau^{\rho\nu}+
\tau^\nu_{ \ d}m^d_{ \ \rho}\tau^{\rho\mu})
+\frac{1}{4\omega^4}\tau^\mu_{ \ a}m_\rho^{ \ a}
h^{\rho\sigma}\tau^\nu_{ \ b}m_\sigma^{\ b} \ . \nonumber \\
\end{eqnarray}
Inserting (\ref{metback}) into (\ref{defGamma}) we obtain in the limit
$\omega \rightarrow \infty$
\begin{eqnarray}\label{Gammalim}
& &\Gamma_{\mu\nu}^{\rho}=
\frac{\omega^2}{2}h^{\mu\sigma}(\partial_\mu \tau_{\sigma\nu}+
\partial_\nu \tau_{\mu\sigma}-\partial_\sigma \tau_{\mu\nu})+\nonumber \\
& &+\frac{1}{2}h^{\mu\sigma}(\partial_\mu \hh_{\sigma\nu}+\partial_\nu \hh_{\sigma\mu}-
\partial_\sigma \hh_{\mu\nu})
+\frac{1}{2}\tau^{\rho\sigma}
(\partial_\mu \tau_{\sigma\nu}+\partial_\nu \tau_{\sigma\mu}-
\partial_\sigma \tau_{\mu\nu})
-\nonumber \\
& &-\frac{1}{4}(\tau^\rho_{ \ b}m_\omega^{ \ }h^{\omega\sigma}+
\tau^\sigma_{ \ b}m_{\omega}^{ \ b}h^{\omega\rho})
(\partial_\mu \tau_{\sigma\nu}+\partial_\nu \tau_{\sigma\mu}-
\partial_\sigma \tau_{\mu\nu}) \ , \nonumber \\
\nonumber \\
\end{eqnarray}
where
\begin{equation}
\hh_{\mu\nu}=
h_{\mu\nu}+\frac{1}{2}\tau_\mu^{ \ a}m_\nu^{ \ b}\eta_{ab}+
\frac{1}{2}m_\mu^{ \ a}\tau_\nu^{ \ b}\eta_{ab} \ . \end{equation}
Let us now focus on the  divergent term  that can be rewritten into the form
\begin{eqnarray}
& &h^{\rho\sigma}(\partial_\mu \tau_{\sigma\nu}+
\partial_\nu \tau_{\mu\sigma}-\partial_\sigma \tau_{\mu\nu})=
\nonumber \\
& &=h^{\rho\sigma}[(\partial_\mu \tau_\sigma^{ \ a}-
\partial_\sigma \tau_\mu^{ \ a})\eta_{ab}\tau_\nu^{ \ b}+
\tau_\mu^{ \ a}(\partial_\nu \tau_\sigma^{ \ b}-\partial_\sigma \tau_\nu^{ \ b})\eta_{ab}] \ .
\nonumber \\
\end{eqnarray}
We see that this term  vanishes when we impose zero torsion
condition
\begin{equation}\label{zerotor}
\partial_\mu \tau_\nu^{ \ a}-\partial_\nu \tau_\mu^{ \ a}=0 \ .
\end{equation}
In what follows we will presume this condition. Then after some calculation
we find that (\ref{Gammalim}) simplifies considerably and has the form
\begin{eqnarray}\label{Gammalimfin}
& &\Gamma_{\mu\nu}^\rho
=
\frac{1}{2}\tau^\rho_{ \ a}(\partial_\mu \tau_\nu^{ \ a}+\partial_\nu \tau_\mu^{ \ a})+
\frac{1}{2}h^{\mu\sigma}(\partial_\mu h_{\sigma\nu}+\partial_\nu h_{\sigma\mu}-
\partial_\sigma h_{\mu\nu})+\nonumber \\
& & +\frac{1}{4}h^{\rho\sigma}((\partial_\mu m_\sigma^{ \ a}-\partial_\sigma m_\mu^{ \ a})\eta_{ab}
\tau_\nu^{ \ b}+(\partial_\nu m_\sigma^{ \ a}-\partial_\sigma m_\nu^{ \ a})\eta_{ab}\tau_\mu^{ \ b}) \ . \nonumber \\
\end{eqnarray}
As a check we can easily find that
 $\tau_\mu^{ \ a}$ and $h^{\mu\nu}$ are covariantly constant
\begin{eqnarray}
\nabla_\mu \tau_\nu^{ \ a}=0 \ , \quad
\nabla_\mu h^{\nu\rho}=0 \ .
\end{eqnarray}
Now we are ready to proceed to the analysis of type IIB supergravity equations
of motion. We start with
 the type IIB string effective action in the form
\begin{equation}
S=\frac{1}{2\tilde{\kappa}_{10}^2}
\int d^{10}x \sqrt{-G_E}\left(R(G_E)-\frac{\partial_\mu \tau G^{\mu\nu}_E
    \partial_\nu\bar{\tau}}{2(Im\tau)^2}-\frac{1}{2}
\frac{|G_3|^2}{Im \tau}\right)\ ,
\end{equation}
where
\begin{equation}
\tau=\chi+i e^{-\Phi}\equiv \tau_R+i\tau_I \ , G^{(3)}=dC^{(2)}-\chi H-ie^{-\Phi}H=
dC_2-\tau H \ ,  \quad H\equiv dB \ ,
\end{equation}
and where the action is written in the Einstein frame where the metric components are related to the string components metric as
\begin{equation}
G^E_{\mu\nu}=e^{-\Phi/2}G_{\mu\nu} \ .
\end{equation}
Finally note that we use the convention
\begin{equation}
|G^{(3)}|^2=\frac{1}{3!}G^{(3)}_{\mu\nu\rho}G^{\mu \mu_1}_E
G^{\nu\nu_1}_E G^{\rho \rho_1}G_{\mu_1\nu_1\rho_1}^{(3)*} \ .
\end{equation}
The variation of the action with respect to $G^{\mu\nu}_E$ gives following equations
of motion for metric
\begin{equation}\label{eqmeff}
R_{\mu\nu}(G_E)-\frac{1}{2}G_{\mu\nu}^ER(G_E)=T_{\mu\nu}
\ ,
\end{equation}
where $T_{\mu\nu}$ is stress energy tensor whose exact definition will be given below. Now taking the trace of the equation (\ref{eqmeff}) with $G^{\mu\nu}_E$ we obtain
\begin{equation}
R(G_E)=-4T \ , T\equiv G^{\mu\nu}_E T_{\mu\nu} \ .
\end{equation}
Inserting this result into (\ref{eqmeff}) we find an alternative form of the equation of motion
\begin{equation}\label{eqGE}
R_{\mu\nu}(G_E)=T_{\mu\nu}-\frac{1}{8}G_{\mu\nu}^E T
\end{equation}
that is suitable for non-relativistic limit. Before we proceed to this point we should analyze one subtle point which is the relation between Einstein frame and string frame connection that generally has the form
\begin{equation}
\Gamma^\rho_{\mu\nu}(G_E)=
\Gamma^\rho_{\mu\nu}(G)-\frac{1}{4}(\delta^\rho_\nu\partial_\mu\Phi+
\delta^\rho_\mu\partial_\nu\Phi-G^{\rho\sigma}\partial_\sigma \Phi \ .
G_{\mu\nu})
\end{equation}
The subtle point is that for the metric  (\ref{metback}) we find that
$\Gamma_{\mu\nu}^\rho(G_E)$ generally diverges in the limit  $\omega \rightarrow \infty$
\begin{eqnarray}
& &\Gamma^\rho_{\mu\nu}(G_E)=
\Gamma^\rho_{\mu\nu}(G)-\frac{1}{4}(\delta^\rho_\nu\partial_\mu\Phi
+\delta^\rho_\mu\partial_\nu\Phi)+\nonumber \\
& &+\frac{1}{4}(\tau^{\rho\sigma}-\frac{1}{2}\tau^\rho_{ \ b}m_\omega^{ \ b}h^{\rho\sigma}-\frac{1}{2}\tau^\sigma_{ \ b}m_\omega^{ \ b}h^{\omega\rho})
\partial_\sigma \Phi\tau_{\mu\nu}+\nonumber \\
& &+\frac{1}{4}h^{\rho\sigma}
\partial_\sigma \Phi (h_{\mu\nu}+\frac{1}{2}\tau_\mu^{ \ a}m_\nu^{ \ b}\eta_{ab}+
\frac{1}{2}m_\mu^{ \ a}\tau_\nu^{ \ b}\eta_{ab})+
\omega^2\frac{1}{4}h^{\rho\sigma}\partial_\sigma\Phi \tau_{\mu\nu} \ .
\nonumber \\
\end{eqnarray}
We see that there is a divergent contribution proportional to $h^{\rho\sigma}\partial_\sigma\Phi$. Then in order to have finite Christoffel symbols in Einstein frame as well we impose following condition on the dilaton
\begin{equation}\label{spatdil}
h^{\mu\nu}\partial_\nu\Phi=0 \ .
\end{equation}
We will discuss this restriction in more details below
\footnote{There is an interesting question whether it is necessary to demand that the Christoffel symbols in string frame should be finite or whether it is not sufficient to demand that they should be finite in Einstein frame only. This requirement would imply relation between derivatives of $\tau_\mu^{ \ a}$ and
    $\partial_\mu\Phi$. We leave this question for further study.}.

Let us now return to the equation of motion (\ref{eqGE}). Since its  left side
 is finite the right side has to be finite too in the limit $\omega\rightarrow \infty$. This fact implies additional conditions  on the background fields. To see this let us focus on the
matter part of the action that depends on $G_{\mu\nu\rho}$ that has following stress energy tensor defined as
\begin{eqnarray}
T_{\mu\nu}^{G_3}=-\frac{1}{\sqrt{-G_E}}\frac{\delta S_{G^{(3)}}}{\delta G^{\mu\nu}_E}=
-\frac{1}{24}G^E_{\mu\nu}G^{(3)}_{\rho\sigma\omega}G^{\rho\rho_1}G^{\sigma\sigma_1}
G^{\omega\omega_1}G^{(3)*}_{\rho_1\sigma_1\omega_1}+
\frac{1}{4}G^{(3)}_{\mu\sigma\omega}G^{\sigma\sigma_1}_EG^{\omega\omega_1}_EG^{(3)*}_{\nu\sigma_1\omega_1} \ .  \nonumber \\
\end{eqnarray}
We see that necessary condition to have finite stress energy tensor
we have to demand that $G^{(3)}_{\mu\nu\rho}$ has to be finite in the limit $\omega\rightarrow \infty$. Let us analyze this requirement in more details and
start for simplicity of notation with the field strength of $B_{\mu\nu}$. Inserting (\ref{ansB}) into  definition of $H_{\mu\nu\rho}$ we obtain
\begin{eqnarray}
& &H_{\rho\mu\nu}=\partial_\rho B_{\mu\nu}+\partial_\mu B_{\nu\rho}+
\partial_\nu B_{\rho \mu}=\nonumber \\
& &=\omega^2\bX[
(\partial_\rho\tau_\mu^{ \ a}-\partial_\mu \tau_\rho^{ \ a})\tau_\nu^{ \ b}\epsilon_{ab}+
(\partial_\nu \tau_\rho^{ \ a}-\partial_\rho \tau_\nu^{ \ a}
)\tau_\mu^{ \ b}\epsilon_{ab}+
(\partial_\mu \tau_\nu^{ \ a}-\partial_\nu \tau_\mu^{ \ a})
\epsilon_{ab}\tau_\rho^{ \ b}]- \nonumber \\
& &-\frac{1}{2}\bX[(\partial_\mu m_\nu^{ \ a}-\partial_\nu m_\mu^{ \ a})
\tau_\rho^{ \ b}\epsilon_{ab}+(\partial_\nu m_\rho^{ \ a}-\partial_\rho
m_\nu^{ \ a})\tau_\mu^{ \ b}\epsilon_{ab}+
(\partial_\rho m_\mu^{ \ a}-\partial_\mu m_\rho^{ \ a})\tau_\nu^{ \ b}
\epsilon_{ab}+\nonumber \\
& & +(\partial_\mu \tau_\nu^{ \ a}-\partial_\nu \tau_\mu^{ \ a})\epsilon_{ab} m_\rho^{ \ b}\epsilon_{ab}+(\partial_\rho\tau_\mu^{ \ a}-\partial_\mu \tau_{\rho}^{ \  a})m_\nu^{ \ b}\epsilon_{ab}+
(\partial_\nu \tau_\rho^{ \ a}-\partial_\rho \tau_\nu^{ \ a})m_\mu^{ \ b}\epsilon_{ab}]
\nonumber \\
& &+\partial_\rho \bX \bB_{\mu\nu}+\partial_\mu \bX \bB_{\nu\rho}+
\partial_\nu \bX \bB_{\rho\mu}+h_{\rho\mu\nu} \ , \quad  h=db \ . \nonumber \\
\end{eqnarray}
First of all we see that the first divergent contribution
proportional to $\omega^2$ vanishes when we impose zero torsion
condition. On the other hand there is still divergent contribution
coming from the terms on the last line that are proportional to
derivative of $\bX$. Since $\bB_{\mu\nu}$ does not vanish and it is
proportional to $\omega^2$ we see that in order to ensure finiteness
of $H_{\rho\mu\nu}$ we have to presume that $\Phi$ and $\chi$ are
constant and hence $\partial_\rho \bX=0$. Note that this is
stronger condition than (\ref{spatdil}). Then however the equations
of motion for $\tau$ implies that $G^{(3)}$ has to be zero. Note
that the condition $G^{(3)}=0$ also solves   the equation of motion
for $G_3$. Further,  the stress energy tensor for complex scalar
$\tau$ vanishes for constant $\tau$. As a result we find that the
equations of motion for stringy NC gravity has simple form
\begin{equation}
R_{\mu\nu}(G_E)=0 \ .
\end{equation}
Thanks to the fact that $\Phi$ is constant this equation also implies
\begin{equation}\label{fineq}
R_{\mu\nu}(G)=0  \ ,
\end{equation}
 where $R_{\mu\nu}$ depends on Christoffel symbols given in (\ref{Gammalimfin}).
\subsection{Special Case $n=0,m=1$}
There is an important special case  when $n=0, m=1$ that corresponds to the situation when the probe that defines stringy Newton-Cartan gravity is fundamental string. This case is special since we have  $\bX=1$. For simplicity we further presume that the background
$RR$ fields are zero. In this case the field strength for $B_{\mu\nu}$ has the form
\begin{eqnarray}\label{Hstring}
& & H_{\rho\mu\nu}=\partial_\rho B_{\mu\nu}+\partial_\mu B_{\nu\rho}+
\partial_\nu B_{\rho \mu}=h_{\mu\nu\rho}-\nonumber \\
& &-\frac{1}{2}[(\partial_\mu m_\nu^{ \ a}-\partial_\nu m_\mu^{ \ a})
\tau_\rho^{ \ b}\epsilon_{ab}+(\partial_\nu m_\rho^{ \ a}-\partial_\rho
m_\nu^{ \ a})\tau_\mu^{ \ b}\epsilon_{ab}+
(\partial_\rho m_\mu^{ \ a}-\partial_\mu m_\rho^{ \ a})\tau_\nu^{ \ b}] \ ,
\nonumber \\
\end{eqnarray}
where we imposed zero torsion condition $\partial_\mu \tau_\nu^{ \ a}-
\partial_\nu \tau_\mu^{ \ a}=0$. Let us again study the equation of motion for the background fields. Note that the relativistic
theory is governed by the action
\begin{equation}
S=\frac{1}{2\tilde{\kappa}_{10}^2}
\int d^{10}x \sqrt{-G_E}[R-\frac{1}{2}\partial_\mu \Phi G^{\mu\nu}_E
    \partial_\nu\Phi-\frac{1}{12}e^{-\Phi}H_{\mu\nu\rho}G^{\mu \mu_1}_E
    G^{\nu\nu_1}_E G^{\rho \rho_1}_EH_{\mu_1\nu_1\rho_1} ] \ ,
\end{equation}
\begin{equation}
G^E_{\mu\nu}=e^{-\Phi/2}G_{\mu\nu} \ .
\end{equation}
Let us start with the equation of motion for $\Phi$
\begin{equation}
\frac{1}{\sqrt{-G_E}}\partial_\mu[\sqrt{-G_E}G^{\mu\nu}_E\partial_\nu \Phi]+
+\frac{1}{12}e^{-\Phi}H_{\mu\nu\rho}G^{\mu \mu_1}_E
G^{\nu\nu_1}_E G^{\rho \rho_1}H_{\mu_1\nu_1\rho_1}=0 \
\end{equation}
and find its non-relativistic limit. First of all we have
\begin{equation}
\det G^E_{\mu\nu}=e^{-5\Phi}\det G_{\mu\nu}=
-e^{-5\Phi}(\det E_\mu^{ \ A})^2 \ .
\end{equation}
Since $E_\mu^{ \ a}=\omega \tau_\mu^{ \ a}+\frac{1}{2\omega}m_\mu^{ \ a} \ ,  E_\mu^{ \ a'}=
e_\mu^{ \ a'}$ we find
\begin{equation}
\det E_\mu^{ \ A}=\omega \det (\tau_\mu^{ \ a},e_\mu^{ \ a'})+
\frac{1}{2\omega}\det (m_\mu^{ \ a},e_\mu^{ \ a'}) \ .
\end{equation}
Then the equation of motion for $\Phi$ reduces into following form in the limit
$\omega\rightarrow \infty$
\begin{equation}\label{eqPhiNC}
\frac{1}{\det (\tau_\mu^{ \ a},e_\mu^{ \ a'})}
\partial_\mu[e^{-2\Phi}\det (\tau_\mu^{ \ a},e_\mu^{ \ a'})h^{\mu\nu}
\partial_\nu \Phi]+
\frac{1}{12}e^{3\Phi}H_{\mu\nu\rho}h^{\mu\mu_1}h^{\nu\nu_1}h^{\rho\rho_1}
H_{\mu_1\nu_1\rho_1}=0\ ,
\end{equation}
where $H_{\mu\nu\rho}$ is given in (\ref{Hstring}).
Further, the equation of motion for $H_{\mu\nu\rho}$ takes the form
\begin{eqnarray}
\partial_\mu[\sqrt{-\det G_E}e^{-\Phi}G^{\mu\mu_1}_EG^{\nu\nu_1}_E
G^{\rho\rho_1}_EH_{\mu_1 \nu_1\rho_1}]=0
\end{eqnarray}
that in the non-relativistic limit reduces into
\begin{equation}\label{eqBNC}
\partial_\mu[ e^{-2\Phi}\det (\tau_\mu^{ \ a},e_\mu^{ \ a'})
h^{\mu\mu_1}h^{\nu\nu_1}h^{\rho\rho_1}H_{\mu_1\nu_1\rho_1}]=0 \ ,
\nonumber \\
\end{equation}
where $H_{\mu\nu\rho}$ is again  given in (\ref{Hstring}).
Let us now discuss consequence of the condition (\ref{spatdil}). Imposing this condition in  (\ref{eqPhiNC}) we obtain following condition on $H_{\mu\nu\rho}$
\begin{equation}\label{Hproject}
h^{\mu\mu_1}h^{\nu\nu_1}h^{\rho\rho_1}H_{\mu_1\nu_1\rho_1}=0 \ .
\end{equation}
Note that in this case the equation of motion  (\ref{eqBNC}) is satisfied as well. Note that in adapted coordinates when $e_\mu^{ \ a'}=e_i^{ \ a'} \ , i=2,\dots,9$ and hence $h^{\mu\nu}$ has non-zero components $h^{ij}$ we find that the condition above implies
\begin{equation}
H_{ijk}=0 \ .
\end{equation}
Let us finally discuss  the equation of motion for $G_{\mu\nu}^E$
 \begin{equation}
 R_{\mu\nu}(G_E)=T_{\mu\nu}-\frac{1}{8}G_{\mu\nu}^E T \ .
 \end{equation}
 As we argued above the left side of this equation is finite in the limit $\omega\rightarrow \infty$.  Then on the right side
 we should have finite contribution as well. In case of the scalar field we obtain
 \begin{equation}
 T_{\mu\nu}^{scal}=-\frac{1}{\sqrt{-\det G_E}}\frac{\delta S^{scal}}{\delta
 G^{\mu\nu}_E}=-\frac{1}{4}G_{\mu\nu}^E G^{\rho\sigma}_E \partial_\rho \Phi\partial_\sigma\Phi+\frac{1}{2}\partial_\mu\Phi\partial_\nu\Phi
\end{equation}
and hence
 \begin{eqnarray}
 T_{\mu\nu}^{scal}-\frac{1}{8}T^{scal}G_{\mu\nu}^E=\frac{1}{2}\partial_\mu\Phi\partial_\nu\Phi \ .
 \nonumber \\
 \end{eqnarray}
On the other hand in case of  $B_{\mu\nu}$ we obtain
 \begin{eqnarray}
 T_{\mu\nu}^B=-\frac{1}{24}G^E_{\mu\nu}H_{\rho\sigma\omega}G^{\rho\rho_1}_EG^{\sigma\sigma_1}_E
 G^{\omega\omega_1}_EH_{\rho_1\sigma_1\omega_1}+
 \frac{1}{4}H_{\mu\sigma\omega}G^{\sigma\sigma_1}_EG^{\omega\omega_1}_EH_{\nu\sigma_1\omega_1} \   \nonumber \\
 \end{eqnarray}
 so that
 \begin{equation}
 T^B=G^{\mu\nu}_ET_{\mu\nu}^B=-\frac{1}{6}H^2 \
 \end{equation}
 which implies
 \begin{equation}
 T_{\mu\nu}^B-\frac{1}{8}G_{\mu\nu}^ET^B=-\frac{1}{48}G^E_{\mu\nu}H^2+
 \frac{1}{4}H_{\mu\rho\sigma}G^{\rho\rho_1}_EG^{\sigma\sigma_1}_EH_{\nu\sigma_1\omega_1} \ .
 \nonumber \\
 \end{equation}
 Note that this expression is finite as well when we impose the condition
 (\ref{Hproject}) however we will not find the explicit result since it is rather complicated.
 As a result we obtain final form of the equation of motion of stringy Newton-Cartan gravity
 \begin{equation}\label{stringyNCg}
 R_{\mu\nu}(G_E)=\frac{1}{2}\partial_\mu\Phi\partial_\nu\Phi+(T_{\mu\nu}^B-\frac{1}{8}G_{\mu\nu}^E T^B)_{finite}\ ,
 \end{equation}
where $\Phi$ has to obey the condition (\ref{spatdil})
that in adapted coordinates  $\tau_\mu^{ \ a}=\delta_\mu^{ \ a},
e_\mu^{ \ a'}=e_i^{ \ a'} \ , i=2,\dots,9$ implies $\partial_i\Phi=0$ and hence
$\Phi=\Phi(t)$. However we should stress one important point which is the fact that the equation of motion for $\Phi$ does not determine its time evolution. The same is true in case of NSNS two form $B_{\mu\nu}$ too. Of course, in both cases we can replace conditions (\ref{spatdil}) and (\ref{Hproject})
stronger conditions $\partial_\mu\Phi=0 \ , H_{\mu\nu\rho}=0$. We leave more general analysis of stringy NC equations of motion for future project.
\section{D1-Brane with Unfixed Gauge Symmetry in NC Background}\label{fifth}
In this section we  consider situation when we have $n$ coincident D1-branes and take the non-relativistic limit on metric and NSNS two form field
that defines stringy NC gravity. We again start with the Hamiltonian given in (\ref{Hgen}) and (\ref{HD1gen}) where  the metric is given in
(\ref{metNCans}) and where NSNS two form field has the form
\begin{eqnarray}
B_{\mu\nu}&=&
(\omega^2\tau_\mu^{ \ a}\tau_\nu^{ \ b}\epsilon_{ab}-
\frac{1}{2}(m_\mu^{ \ a}\tau_{\nu}^{ \ b}+
\tau_\mu^{\  a}m_\nu^{ \ b})\epsilon_{ab}+\frac{1}{4\omega^2}
m_\mu^{ \ a}m_\mu^{ \ b}\epsilon_{ab})+b_{\mu\nu} \ . \nonumber \\
\end{eqnarray}
With such a choice of the NSNS field we find that divergent contribution in the Hamiltonian constraint that are proportional to  $\tau_{\mu\nu}$ cancel for any $\pi^\sigma$.
However there are now several possibilities how to take scaling limit of RR fields. For example, if we demand
that $C^{(2)}_{\mu\nu}$ has similar form as  $B_{\mu\nu}$, i.e. $C^{(2)}_{\mu\nu}=
\bB_{\mu\nu}+c_{\mu\nu}$ the requirement that   Hamiltonian
constraint should be finite in the limit $\omega\rightarrow \infty$ implies following condition
\begin{equation}
2\pi^\sigma n=-2\pi^\sigma n \chi  \ , \quad
e^{-2\Phi}+\chi^2=1 \ .
\end{equation}
The first equation implies $\chi=-1$ that together with the second
one gives  $e^{-2\Phi}=0$. However this is the realm of the infinite coupled
string theory which is clearly rather  subtle
limit. For that reason we consider another scaling limit when
\begin{equation}
n=\frac{1}{\omega}\tn \ , \quad \chi=\frac{1}{\omega}\tchi \  , \quad  C^{(2)}_{\mu\nu}=\omega\tC_{\mu\nu}
\ .
\end{equation}
Now the Hamiltonian constraint is finite and has the form
\begin{eqnarray}
& &\mH_\tau=\Pi_\mu h^{\mu\nu}\Pi_\nu-2T_{D1}\pi^\sigma \Pi_\mu \htau^\mu_{ \ a}
\eta^{ab}\epsilon_{bc}\tau_\sigma^{ \ c}-
T_{D1}^2 (\pi^\sigma)^2 \tau_\sigma^{ \ a}\epsilon_{ab}\Phi^{bc}\epsilon_{cd}
\tau_\sigma^{ \ d}+
\nonumber \\
& &+T_{D1}^2 (\pi^\sigma)^2 \partial_\sigma x^\mu \bar{h}_{\mu\nu}\partial_\sigma x^\nu+
T_{D1}^2 e^{-2\Phi}\tn^2\partial_\sigma x^\mu\tau_{\mu\nu}\partial_\sigma x^\nu \ ,
\nonumber \\
\end{eqnarray}
where
\begin{eqnarray}
\Pi_\mu=p_\mu-T_{D1}(\pi^\sigma b_{\mu\rho}+\tn \tC_{\mu\rho})
\partial_\sigma x^\rho \ , \nonumber \\
\bar{h}_{\mu\nu}=h_{\mu\nu}+m_\mu^{ \ a}\tau_\nu^{ \ b}\eta_{ab}+
\tau_\mu^{ \ a}m_\nu^{ \ b}\eta_{ab} \ .
\nonumber \\
\end{eqnarray}
Note that $\pi^\sigma$ is still dynamical field which is a crucial
difference when we compare with the analysis presented in  section (
\ref{third}).

As the next step we determine corresponding Lagrangian. From the Hamiltonian
\begin{equation}
H=\int d\sigma (N^\tau \mH_\tau+N^\sigma \mH_\sigma+\partial_\sigma A_\tau\pi^\sigma)
\end{equation}
we obtain
\begin{eqnarray}
& &\partial_\tau x^\mu=\pb{x^\mu,H}=
2N^\tau h^{\mu\nu}\Pi_\nu-2N^\tau T_{D1}\pi^\sigma
\htau^\mu_{ \ a}\eta^{ab}\epsilon_{bc}\tau_\sigma^{ \ c}+N^\sigma \partial_\sigma x^\mu \ , \nonumber \\
& &\partial_\tau A_\sigma=\pb{A_\sigma,H}=\partial_\sigma A_\tau
-2N^\tau T_{D1}b_{\mu\rho}\partial_\sigma x^\rho h^{\mu\nu}\Pi_\nu
-2N^\tau T_{D1}\Pi_\mu \htau^\mu_{ \ a}\eta^{ab}\epsilon_{bc}\tau_\sigma^{ \ c}+
\nonumber \\
& &+2N^\tau T_{D1}^2 b_{\mu\rho}\partial_\sigma x^\rho \pi^\sigma \htau^\mu_{ \ a}\eta^{ab}
\epsilon_{bc}\tau_\sigma^{ \ c}-2N^\tau T_{D1}^2\pi^\sigma \tau_\sigma^{ \ a}\epsilon_{ab}
\Phi^{bc}\epsilon_{cd}\tau_\sigma^{ \ d}+\nonumber \\
& &+2N^\tau T_{D1}^2\pi^\sigma \partial_\sigma x^\mu\bar{h}_{\mu\nu}\partial_\sigma x^\nu \ .
\nonumber \\
\end{eqnarray}
Then the Lagrangian density has the form
\begin{eqnarray}
& &\mL
=N^\tau \Pi_\mu  h^{\mu\nu}\Pi_\nu
+T_{D1} \tn X^\mu \tC_{\mu\nu}\partial_\sigma x^\nu-
2N^\tau \pi^\sigma T_{D1}\Pi_\mu \htau^{\mu}_{ \ a}\eta^{ab}\epsilon_{bc}\tau_\sigma^{ \ c}+\nonumber \\
& & +N^\tau T_{D1}^2(\pi^\sigma)^2 \tau_\sigma^{ \ a}
\epsilon_{ab}\Phi^{bc}\epsilon_{cd}\tau_\sigma^{ \ d}
-N^\tau T_{D1}^2 (\pi^\sigma)^2 \partial_\sigma x^\mu
\bh_{\mu\nu}\partial_\sigma x^\nu-e^{-2\Phi}T_{D1}^2N^\tau \tn^2\ba_{\sigma\sigma} \ .
\nonumber \\
\end{eqnarray}
To proceed further we use the fact that
\begin{equation}
\he_\mu^{ \ a'}\tau^\mu_{ \ b}=0
\end{equation}
and we again obtain
\begin{equation}
X^\mu \he_\mu^{ \ a'}\delta_{a'b'}\he_\nu^{ \ b'}X^\nu=4(N^\tau)^2\Pi_\mu h^{\mu\nu}\Pi_\nu \ .
\end{equation}
As the next step  we express $\htau^\mu_{ \ a}\Pi_\mu$ with the help of the equation of motion for $A_\sigma$ as
\begin{eqnarray}
& &F_{\tau\sigma}+T_{D1}X^\mu b_{\mu\nu}\partial_\sigma x^\nu+
T_{D1}\frac{\sqrt{-\det\ba}}{\ba_{\sigma\sigma}}
\tau_\sigma^{ \ a}\epsilon_{ab}\Phi^{bc}\epsilon_{cd}\tau_\sigma^{\  d}
-T_{D1}\frac{\sqrt{-\det\ba}}{\ba_{\sigma\sigma}}\partial_\sigma x^\mu
\bh_{\mu\nu}\partial_\sigma x^\nu=
\nonumber \\
& & =-2N^\tau T_{D1}\Pi_\mu\htau^\mu_{ \ a}\eta^{ab}\epsilon_{bc}\tau_\sigma^{ \ c} \ ,
\nonumber \\
\end{eqnarray}
where we also used the fact equation of motion for $x^\mu$ implies
\begin{equation}\label{Nsol5}
N^\sigma=\frac{\ba_{\tau\sigma}}{\ba_{\sigma\sigma}} \ , \quad
\pi^\sigma N^\tau=\frac{\sqrt{-\det\ba}}{2 \ba_{\sigma\sigma}T_{D1}} \
\end{equation}
and hence we expressed $\pi^\sigma$ with the help of the second equation in
(\ref{Nsol5}).
Collecting all these terms together and after some calculations we obtain
Lagrangian density in the form
\begin{eqnarray}\label{mLstringy}
& &\mL=\frac{\det\ba}{4N^\tau\ba_{\sigma\sigma}}\left(\ba^{\alpha\beta}\bh_{\alpha\beta}+
\ba^{\alpha\beta}\tau_\alpha^{ \ a}\eta_{ab}\Phi^{bc}\eta_{cd}\tau_\beta^{ \ d}-\Phi^{ab}\eta_{ba}-
\frac{2}{\sqrt{-\det\ba}}(F_{\tau\sigma}+T_{D1}b_{\tau\sigma})\right)-\nonumber \\
& & -e^{-2\Phi}N^\tau T_{D1}^2 \tn^2\ba_{\sigma\sigma}+T_{D1}\tn \tC_{\tau\sigma} \ .
\nonumber \\
\end{eqnarray}
Finally solving equation of motion for $N^\tau$ and inserting the result into (\ref{mLstringy}) we obtain final form of the Lagrangian density for D1-brane in stringy NC background
\begin{equation}
\mL=-\tn e^{-\Phi}T_{D1}\sqrt{-\det\ba}
\sqrt{\ba^{\alpha\beta}\bh_{\alpha\beta}-\frac{1}{\sqrt{-\det\ba}}\epsilon^{\alpha\beta}
    (F_{\alpha\beta}+T_{D1}b_{\alpha\beta})}+T_{D1}\tn \tC_{\tau\sigma} \ .
\end{equation}
This is the Lagrangian density for $\tn-$ D1-branes in stringy NC background.

Analysis of the consistency of the background fields is the same as in section (\ref{fourth}) since we consider stringy NC background. On the other hand since RR two form $C^{(2)}$ scales as $\omega$ we should demand that its field strength vanishes in order to have finite stress energy tensor. Then clearly the equations of motion for $C^{(2)}$ are obeyed as well.
\section{Conclusion}\label{sixth}
Let us outline results derived in this paper and suggest possible extension of this work. The main goal of this paper was to analyze properties of D1-brane theory in the limit that defines stringy NC background from relativistic theory
\cite{Bergshoeff:2015uaa}. We firstly considered D1-brane theory with fixed gauge invariance that defines $(m,n)-$ string and we found Lagrangian density for $(m,n)-$string in stringy NC background.  Then we studied conditions that the background fields have to obey in order to define consistent theory. We also found that the  $(m,n)-$string in stringy NC background  is manifestly $SL(2,Z)$ invariant as a consequence of $SL(2,Z)$ invariance of the parent type IIB theory.

We further  analyzed the problem of general D1-brane in stringy NC
background. We showed that in order to have non-trivial D1-brane
theory we should scale the number of D1-branes and RR two form in an
appropriate way. We find corresponding Hamiltonian for D1-brane with
dynamical gauge field and corresponding Lagrangian density that is
manifestly gauge invariant. We mean that this is nice result that
brings new insight into the definition of the action for  D1-branes
in stringy NC gravity.

This work can be extended in several directions. For example, it would be nice to analyze limit when the string coupling goes to infinity in more details. Further, we would like to
analyze T-duality along world-volume direction of D1-brane in NC theory. With analogy with standard relativistic string theory we should expect that resulting object is D0-brane in T-dual theory. It would be nice to analyze similar situation in stringy NC theory in more details. It would be also very nice to better understand to non-relativistic limit of the equations of motion of type IIB gravity. In particular, it is not clear whether it is necessary to impose condition that Christoffel symbols should be finite in string frame or whether it is sufficient to demand that they are finite in Einstein frame only. It would be also very nice to find relation between limit studied in this paper and recent interesting results derived in
\cite{Hansen:2018ofj,Bergshoeff:2018vfn}
  We hope to return to this problem in future.

%


\end{document}